\documentclass[10pt,a4paper]{article}

\usepackage[T1]{fontenc}
\usepackage[utf8]{inputenc}
\usepackage[a4paper,top=1.55cm,bottom=1.7cm,left=1.55cm,right=1.55cm]{geometry}
\usepackage{newtxtext,newtxmath}
\usepackage{microtype}
\usepackage{graphicx}
\usepackage{amsmath,mathtools,bm}
\usepackage{booktabs,tabularx,array}
\usepackage{siunitx}
\usepackage{xcolor}
\usepackage{ragged2e}
\usepackage{titlesec}
\usepackage[font=small,labelfont=bf]{caption}
\usepackage[numbers,sort&compress,super]{natbib}
\usepackage[colorlinks=true,linkcolor=blue,citecolor=blue,urlcolor=blue]{hyperref}
\usepackage{dblfloatfix}
\usepackage{balance}
\usepackage[switch,mathlines]{lineno}
\modulolinenumbers[1]

\allowdisplaybreaks[2]
\titleformat{\section}
  {\large\sffamily\bfseries}
  {}{0pt}{}
\titleformat{\subsection}
  {\normalsize\sffamily\bfseries}
  {}{0pt}{}
\titlespacing*{\section}{0pt}{1.1ex plus .3ex minus .2ex}{0.55ex}
\titlespacing*{\subsection}{0pt}{0.9ex plus .3ex minus .2ex}{0.35ex}

\DeclareCaptionLabelSeparator{naturebar}{\enspace|\enspace}
\newcommand{\safeincludegraphics}[2][]{%
  \IfFileExists{#2}{\includegraphics[#1]{#2}}{%
    \fbox{\parbox[c][36mm][c]{0.92\linewidth}{\centering
    \textsf{Figure file not found:}\\[2pt]\texttt{#2}}}}%
}

\begin{document}
\twocolumn[
\begin{minipage}{\textwidth}
\centering
{\LARGE\sffamily\bfseries Fluctuation-Driven Nonlinear Amplification of Quantum Statistics\par}
\vspace{0.8em}
{\normalsize Yuewei Song$^{1,2,\dagger}$, Zhenghe Zhou$^{1,2,\dagger}$, Shuai Wan$^{1,2,\dagger}$, Hecheng Wang$^{1,2}$, Jinpeng Li$^{1,2}$, Bowen Liu$^{1,2}$, Yinhai Li$^{1,2}$, Chunhua Dong$^{1,2}$, Guangcan Guo$^{1,2}$, Chong Wang$^{3,*}$, Zhiyuan Zhou$^{1,2,*}$, Baosen Shi$^{1,2,*}$\par}
\vspace{0.55em}
{\footnotesize $^{1}$Laboratory of Quantum Information, University of Science and Technology of China, Hefei, Anhui 230026, China\\
$^{2}$CAS Center for Excellence in Quantum Information and Quantum Physics, University of Science and Technology of China, Hefei 230026, China\\
$^{3}$School of Earth and Space Science, University of Science and Technology of China, Hefei 230026, China\\
$^{\dagger}$These authors contributed equally to this work.\\
$^{*}$Correspondence:  \href{mailto:wclhy50@ustc.edu.cn}{wclhy50@ustc.edu.cn} (C.W.); \href{mailto:zyzhouphy@ustc.edu.cn}{zyzhouphy@ustc.edu.cn} (Z.Y.Z.); \href{mailto:drshi@ustc.edu.cn}{drshi@ustc.edu.cn} (B.S.S.)\par}
\vspace{0.85em}
\begin{minipage}{0.94\textwidth}
\small
\justifying
\setlength{\parindent}{0pt}
Photon statistics have moved to the forefront of modern optics, as intensity fluctuations and correlations shape multiphoton interactions and reveal information beyond mean-intensity measurements. Developing high-quality photon sources with pronounced correlations is a fundamental necessity in these fields. Here we demonstrate fluctuation-driven nonlinear statistical amplification of quantum light in spontaneous four-wave mixing using filtered amplified spontaneous emission (ASE). Extending the coherent-pump framework to fluctuating fields, we show how nonlinear weighting of pump intensity combines with bosonic bunching to amplify quantum statistics and reshape temporal correlations. In a SiN microring, ASE pumping increases the zero-delay unconditional second-order correlation from 2.01 to 7.58 and extends the Hanbury Brown--Twiss correlation time by a factor of approximately 2.4. The super-bunched quantum source nevertheless retains heralded single-photon behaviour with $g_H^{(2)}(0)\simeq0.04$, while the same ASE pump supports time--energy entanglement in a silicon waveguide with a raw Franson visibility of 89.84\%. These results establish driving-field statistics as a design dimension for quantum light, broadening the horizons for research into higher-order correlations and nonlinear physics.
\end{minipage}
\end{minipage}
\vspace{1.1em}
]
\section*{Introduction}\label{sec:introduction}

Nonlinear optical interactions lie at the heart of modern photonics, shaping phenomena that range from ultrafast and strong-field dynamics to the generation and engineering of quantum states of light.\cite{dutt2024,wang2020,pelucchi2022,caspani2017,chang2014,lewenstein2021} In multiphoton processes, the optical response is governed not only by the average intensity but also by the photon statistics of the driving field.\cite{lamprou2020} Recent experiments have demonstrated that quantum-statistical fluctuations can substantially enhance nonlinear processes.\cite{spasibko2017,gorlach2023,rasputnyi2024,heimerl2024,wu2026,lemieux2025,gorlach2020,pizzi2023} Photon statistics have consequently emerged as a physical resource for controlling light--matter interactions.\cite{stammer2023,you2026,liu2025,heimerl2025} Intensity correlations in fluctuating light also enable information to be retrieved beyond conventional direct measurements\cite{lubin2022,tsao2025}, as exemplified by ghost imaging and super-bunching-based ranging.\cite{moreau2019,pearce2026,gatti2004,staffas2026,zhang2026lidar,yan2026,zhou2017,dong2024,jahnke2016,lemieux2019,hartmann2015} Together, these advances have brought photon statistics to the forefront of modern optics and stimulated growing interest in light with pronounced intensity fluctuations.\cite{dorfman2020,schlawin2018,dayan2004,you2023} Accessing and exploiting such statistics is becoming increasingly important for statistics-sensitive processes.\cite{casalengua2020} Yet the statistical properties of optical fields remain largely dictated by their generation mechanisms, leaving limited scope for reshaping them independently. Existing studies have focused primarily on generating or exploiting particular statistical states.\cite{zhou2017,spasibko2017,gatti2004}  The nonlinear creation and evolution of higher-order correlations during quantum--state generation remain underexplored.

In quantum optics, the role of driving-field statistics in spontaneous nonlinear processes has received less attention, as these processes are conventionally driven by coherent lasers with weak intensity fluctuations. Consequently, the statistics of the generated field are primarily attributed to the mechanism of nonlinear interactions itself. In this scenario, spontaneous-emission sources provide a natural way to investigate this issue due to their intrinsic super-Poissonian distribution. Multiphoton generation depends nonlinearly on the instantaneous optical intensity. The resulting preferential weighting of high-intensity events can substantially reshape the photon-number distribution of the resulting field. Thermal-like pumping therefore opens a route to exploring the evolution of statistics during quantum--state generation. Recent studies have shown that low-coherence and incoherent light can support high-quality nonlinear and quantum optical processes.\cite{hutter2020,zhang2023led,li2023,song2025,li2026sun,song2026hom,xing2026,jha2010,ismail2017,zhang2019} Among these sources, amplified spontaneous emission (ASE) is particularly advantageous. Investigating statistical properties poses a practical challenge: the broad optical bandwidth typical of spontaneous emission results in extremely short coherence times. Finite temporal resolution averages over multiple temporal modes and suppresses the observable bunching. Spectral filtering can extend the coherence time and make thermal-like fluctuations experimentally accessible, although at the cost of optical power.\cite{pietralunga2003,hartmann2017} ASE can be efficiently amplified and transmitted in a single spatial mode, providing sufficient spectral power density even after narrowband filtering. The resulting field retains enough power to drive nonlinear response while exhibiting pronounced bunching on a resolvable temporal scale. Filtered ASE thus provides an experimental platform for examining the evolution of pump statistics through a spontaneous multiphoton quantum process.

\begin{figure*}[t]
    \centering
    \safeincludegraphics[width=1.0\textwidth]{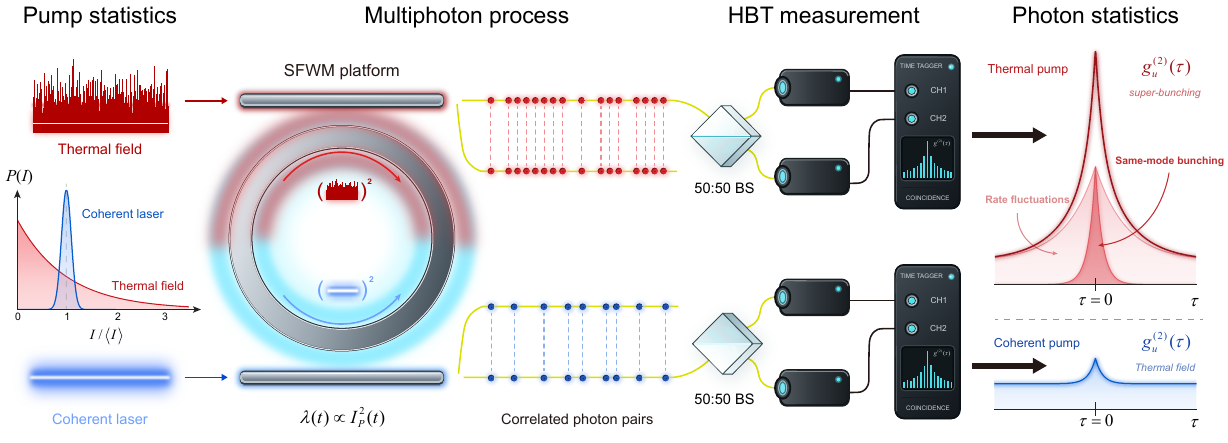}
    \caption{\textbf{Statistical amplification in spontaneous four-wave mixing.}
    Thermal-like intensity fluctuations of the pump are weighted by the quadratic SFWM interaction, for which the instantaneous pair-generation rate scales as $I_p^2(t)$. The resulting HBT correlation contains fluctuations of the pair-generation rate together with the bosonic bunching of the generated photons. Coherent pumping provides the conventional thermal photon-statistics reference.}
    \label{fig:mechanism}
\end{figure*}

Here, we report fluctuation-driven nonlinear amplification of quantum statistics in spontaneous four-wave mixing (SFWM) driven by filtered ASE. The nonlinear dependence of SFWM on pump intensity gives disproportionate weight to high-intensity excursions, producing substantially stronger photon-number fluctuations in the generated field. Extending the coherent-pump framework to fluctuating fields, we identify the combined contributions of pump-induced fluctuations in the pair generation rate and bosonic bunching. Together with the cavity response, these contributions determine both the peak value and temporal profile of the resulting super-bunching. In a SiN microring, ASE pumping increases the unconditional second-order correlation from 2.01 under coherent pumping to 7.58, nearly four times the single-mode thermal value. The Hanbury Brown--Twiss (HBT) correlation time simultaneously increases by a factor of approximately 2.4, revealing an additional temporal scale inherited from the pump. The driving field thus shapes photon correlations beyond the timescale set by the cavity photon lifetime. The statistical amplification remains compatible with the essential quantum properties of the generated states. The super-bunched quantum state shows excellent heralded single-photon behaviour with $g_H^{(2)}(0)\simeq0.04$, while a complementary experiment in a silicon straight waveguide demonstrates high-quality time--energy entanglement under the same ASE pump. These results establish a direct connection between pump fluctuations and the statistics of spontaneously generated quantum light. By placing photon statistics within the physics of nonlinear quantum state generation, this work enriches the conventional coherent-pumping framework. It offers a broader strategy for controlling multiphoton processes, with potential implications extending beyond quantum photonics to fluctuation-sensitive nonlinear physics at large.

\section*{Results}\label{sec:results}

\subsection*{Statistical amplification in SFWM}

Pump-intensity fluctuations influence spontaneous quantum light generation through the nonlinear weighting of different intensity levels. In a multiphoton interaction, the generation probability depends on higher-order moments of the pump intensity. Large intensity excursions acquire disproportionately greater weight through the nonlinear coupling. Fig.~\ref{fig:mechanism} illustrates how fluctuations of the driving field can consequently develop into much stronger fluctuations of the generated photons. Spontaneous four-wave mixing (SFWM) provides a natural setting for this transformation because each pair creation event involves the annihilation of two pump photons. Treating the pump as a classical field with slowly varying amplitude $\alpha_p(t)$, the interaction between the selected signal and idler modes is described by\cite{helt2010,vernon2015lossy}

\begin{equation}
\begin{aligned}
\hat H_{\mathrm{int}}(t)
&=
i\hbar\left[
\mathcal G(t)\hat a_s^\dagger\hat a_i^\dagger
-
\mathcal G^*(t)\hat a_s\hat a_i
\right],\\
\mathcal G(t)&=g\alpha_p^2(t).
\end{aligned}
\label{eq:sfwm-hamiltonian}
\end{equation}
where $g$ is the effective SFWM coupling coefficient, incorporating the third-order nonlinear susceptibility and the spatial overlap. The pair creation amplitude follows the square of the pump field, while the instantaneous generation strength scales as $|\mathcal G(t)|^2\propto I_p^2(t)$.

Resolving the intensity fluctuations of filtered ASE requires a coherence time long enough for temporal correlation measurements. Resonant SFWM is well suited to this requirement. It provides narrow-linewidth photon pairs with high spectral brightness, while the photon lifetime is long enough to resolve the temporal correlation. In our system, the ASE fluctuations remain relevant over the finite response time of the microring, so the generated photon flux cannot be described solely by the pump intensity at the observation time. An excitation of cavity mode $\mu=s,i$ at time zero leaves a field amplitude that decays as $e^{-\kappa_\mu t/2}$ and accumulates a detuning-dependent phase $e^{-i\Delta_\mu t}$. This causal cavity-amplitude response is $h_\mu(t)=\exp[-(\kappa_\mu/2+i\Delta_\mu)t]\Theta(t)$, where $\kappa_\mu$ is the energy-decay rate and $\Theta(t)$ sets the response to zero before the excitation.\cite{vernon2015lossy,helt2010} In the spontaneous low-gain limit, the perturbative solution contains the response
\begin{equation}
\mathcal K(t,v;\mathcal G)
=
\int_v^t
h_s(t-u)\,
\mathcal G(u)\,
h_i^*(u-v)\,
\mathrm du .
\label{eq:causal-response}
\end{equation}

\begin{figure}[t]
    \centering
    \safeincludegraphics[width=0.98\columnwidth]{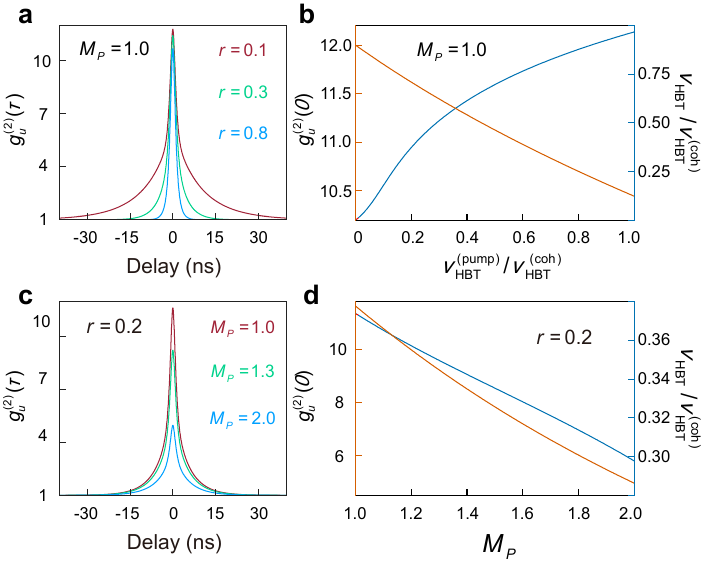}
    \caption{\textbf{Dependence of statistical amplification on pump fluctuations and temporal coherence.}
    \textbf{a,} Calculated HBT correlations for an ideal single-mode thermal pump ($M_p=1$) at different pump-to-photon bandwidth ratios $r_\nu$.
    \textbf{b,} Zero-delay unconditional correlation $g_{u}^{(2)}(0)$ and normalized effective HBT linewidth as functions of $r_\nu$.
    \textbf{c,} HBT correlations at fixed $r_\nu=0.2$ for different effective pump mode numbers $M_p$.
    \textbf{d,} Corresponding $g_{u}^{(2)}(0)$ and normalized HBT linewidth as functions of $M_p$. The generated photon field is taken to be single mode in all calculations.}
    \label{fig:theory}
\end{figure}

\begin{figure*}[t]
    \centering
    \safeincludegraphics[width=0.97\textwidth]{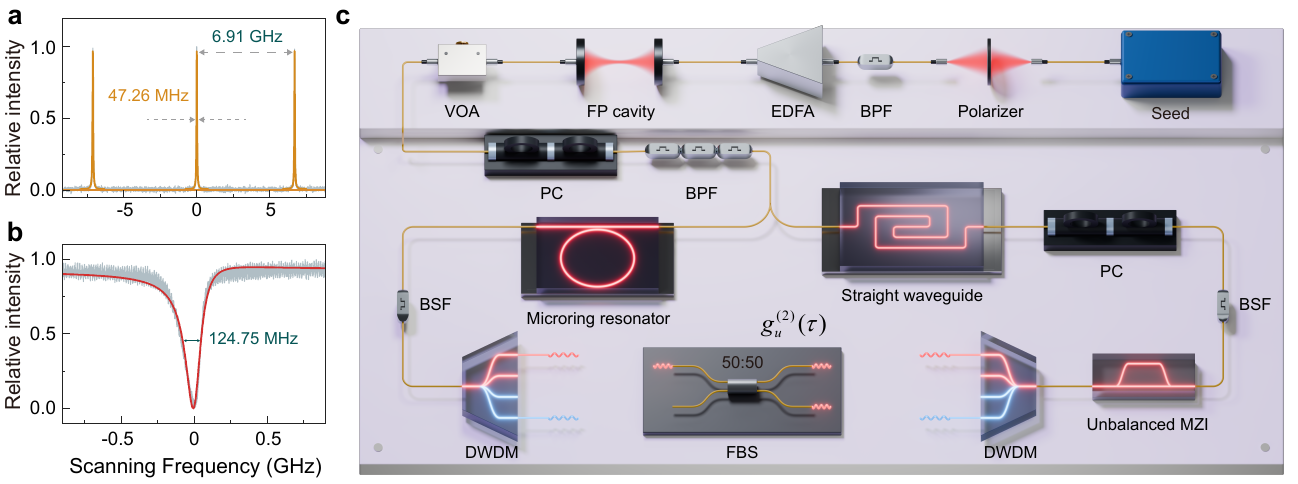}
    \caption{\textbf{Experimental platform.}
    \textbf{a,} Transmission spectrum of the final FP filter. 
    \textbf{b,} Transmission spectrum of the microring resonance . 
    \textbf{c,} Experimental setup for pump preparation and the microring and waveguide measurements.
    Filtered ASE and a narrow-linewidth coherent laser are used as the pump fields. The ASE is spectrally prepared by bandpass filtering, optical amplification and narrowband Fabry--Pérot (FP) filtering before being coupled to either the SiN microring or the silicon strip waveguide. Representative spectra show the final FP filter (FSR, $6.91$ GHz; scanned linewidth, $47.26$ MHz) and the microring resonance (linewidth, $124.75$ MHz). The asymmetric microring transmission feature is described by an empirical asymmetric fit. The microring branch is used for HBT, signal--idler correlation and heralded measurements. The waveguide branch includes an unbalanced Mach--Zehnder interferometer for the time--energy interference measurements. BPF, bandpass filter; VOA, variable optical attenuator; PC, polarization controller; BSF, band-stop filter; DWDM, dense wavelength-division multiplexer; FBS, fibre beam splitter.}
    \label{fig:setup}
\end{figure*}

In Eq.~\eqref{eq:causal-response}, $v$ denotes the time at which the idler vacuum input enters the response. The nonlinear interaction occurs at $u$, and the generated signal is observed at $t$. The pump amplitude is weighted first by the idler response and then by the signal response. The detected signal thus depends on the pump field over a time interval set by the cavity decay. Its statistics depend on both the fluctuation strength and the pump coherence time relative to the photon lifetime. The full Heisenberg--Langevin treatment and the derivation of Eq.~\eqref{eq:causal-response} are given in Methods and Supplementary Materials.

Bunching beyond the conventional thermal value can also be explained within this framework. For a given pump field $\alpha_p(t)$, the low-gain signal field is a zero-mean Gaussian state. Its fourth-order moments can then be factorized into intensity and exchange terms. We denote the first-order coherence by $C_{\mathcal G}(t,t')=\langle\hat b_{se,\mathrm{out}}^\dagger(t)\hat b_{se,\mathrm{out}}(t')\rangle_{\mathcal G}$ for the monitored signal output field $\hat b_{se,\mathrm{out}}$, and define the corresponding photon flux as $N_{\mathcal G}(t)=C_{\mathcal G}(t,t)$. Averaging the Gaussian fourth-order correlation over the pump fluctuations gives the unconditional signal-field second-order correlation

\begin{equation}
\begin{aligned}
g_{u}^{(2)}(\tau)
&=
\frac{
\left\langle
N_{\mathcal G}(t)N_{\mathcal G}(t+\tau)
\right\rangle_{\mathcal G}
+
\left\langle
\left|C_{\mathcal G}(t,t+\tau)\right|^2
\right\rangle_{\mathcal G}
}{\bar N^2}\\
&\equiv
B_N(\tau)+X_C(\tau),
\qquad
\bar N=\langle N_{\mathcal G}\rangle_{\mathcal G}.
\end{aligned}
\label{eq:hbt-decomposition}
\end{equation}

The two contributions have different physical origins. $B_N(\tau)$ describes correlations of the photon flux after the resonator response. $X_C(\tau)=\langle|C_{\mathcal G}(t,t+\tau)|^2\rangle_{\mathcal G}/\bar N^2$ is the bosonic exchange term from Gaussian moment factorization. It contributes when the two detection events are indistinguishable within the same detected optical mode. For one detected mode without pump-driven rate fluctuations, this term gives thermal statistics with $g_{u}^{(2)}(0)=2$.

When the pump intensity fluctuates, the pair generation rate changes with it. At zero delay, $C_{\mathcal G}(t,t)=N_{\mathcal G}(t)$, so one detected mode gives $g_{u}^{(2)}(0)=2B_N(0)=2[1+\mathrm{Var}_{\mathcal G}(N_{\mathcal G})/\bar N^2]$. Here, $\mathrm{Var}_{\mathcal G}$ denotes the variance over the fluctuating pump field. The factor of two is the thermal bunching of the generated signal field. The additional factor $B_N(0)>1$ arises from fluctuations of the pair flux. Their temporal dependences are also different. $B_N(\tau)$ carries the timescale of the pump-driven rate fluctuations after the resonator response. $X_C(\tau)$ is governed mainly by the temporal overlap of the detected optical mode. The measured HBT trace can consequently extend beyond the photon response timescale inferred from coherent pumping.

The nonlinear amplification is most transparent when the pump varies slowly compared with the photon lifetime. For one detected optical mode, this quasi-static limit gives $N_{\mathcal G}\propto|\mathcal G|^2\propto I_p^2$ and $g_{u}^{(2)}(0)=2\langle I_p^4\rangle/\langle I_p^2\rangle^2$. A single thermal pump mode then gives $g_u^{(2)}(0)=12$. The unconditional correlation is governed by higher-order moments of the pump intensity. This relation explains how nonlinear weighting can strongly amplify modest pump bunching. See more details in Supplementary Materials.

For the calculations in Fig.~\ref{fig:theory}, we represent the pump by two statistically independent thermal modes with the same temporal correlation function (see Methods and Supplementary Materials for more details). SFWM annihilates two pump photons, which may come from the same mode or one from each mode. The corresponding generated intensities are added incoherently after propagation through the full signal and idler cavity responses. The relative pump-mode weights determine the effective mode number $1\le M_p\le2$. The temporal scales are specified by the pump HBT-equivalent linewidth $\nu_P$ and the coherent-pump photon HBT linewidth $\nu_Q$, with $r_\nu=\nu_P/\nu_Q$.

\begin{figure*}[t]
    \centering
    \safeincludegraphics[width=0.97\textwidth]{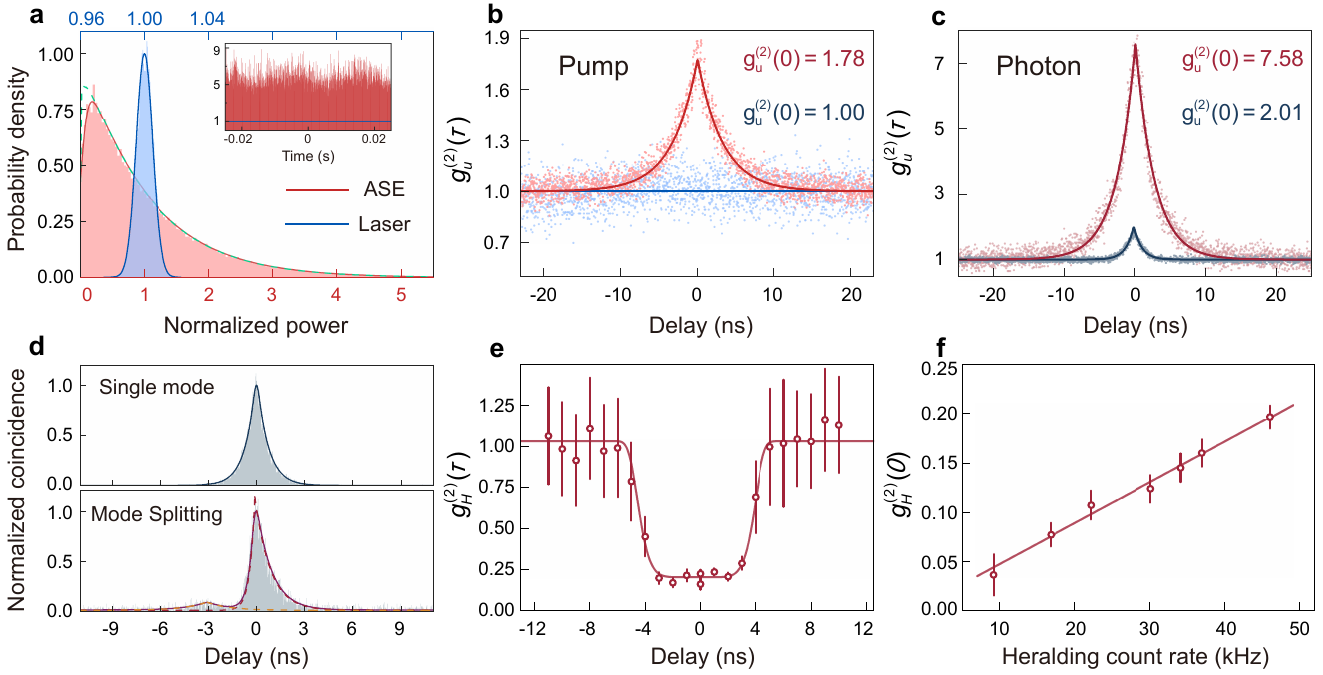}
    \caption{\textbf{Fluctuation-driven statistical amplification in the SiN microring.}
    \textbf{a,} Normalized intensity distributions of the filtered ASE pump and coherent reference. The dashed curve shows the inferred thermal intensity distribution obtained from the noise-broadened model; the inset shows a representative ASE temporal trace.
    \textbf{b,} Pump HBT correlations, giving $g_{u}^{(2)}(0)=1.78$ for ASE and $1.00$ for the coherent laser.
    \textbf{c,} Unconditional photon HBT correlations, with $g_{u}^{(2)}(0)=7.58\pm0.04$ under ASE pumping and $2.01$ under coherent pumping.
    \textbf{d,} Signal--idler correlations under coherent and ASE pumping. A two-component fit to the ASE-pumped trace gives an effective $M_q=1.31\pm0.07$ when the components are treated as mutually incoherent detected modes.
    \textbf{e,} Heralded HBT correlation under ASE pumping at a heralding count rate of approximately $46$ kHz, with $g_H^{(2)}(0)\simeq0.2$.
    \textbf{f,} Heralded $g_H^{(2)}(0)$ as a function of heralding count rate.}
    \label{fig:ring}
\end{figure*}

 Fig.~\ref{fig:theory} show the result for a single thermal pump mode as $r_\nu$ is varied. Increasing $r_\nu$ shortens the pump fluctuation time relative to the photon response and suppresses the bunching enhancement. Figs.~\ref{fig:theory}c,d show the effect of increasing $M_p$ at fixed $r_\nu=0.2$. The peak decreases because increasing $M_p$ weakens the pump-driven rate fluctuations and reduces bosonic bunching by distributing the generated intensity more evenly among optical modes. In this scan, the exchange contribution, which decays on the shorter photon-response timescale, decreases more strongly than the slower rate-correlation contribution. The slower fluctuations therefore account for a larger fraction of the remaining correlation, extending its duration and reducing the effective HBT linewidth.

\subsection*{Microring realization}

The preceding analysis requires appreciable pump fluctuations that remain correlated over an experimentally accessible timescale. Broadband ASE has thermal field statistics, but its short coherence time causes many temporal modes to be averaged within a practical measurement window. The observable bunching is consequently strongly suppressed. We narrow the ASE bandwidth before nonlinear generation to extend the fluctuation timescale. Optical amplification compensates for the accompanying loss of spectral power. Under linear phase-insensitive amplification, a zero-mean Gaussian thermal field retains its thermal character when the selected spectral modes are not substantially reshaped. The corresponding derivation is given in Supplementary Materials. In our experiment, the amplified ASE is successively filtered to a final HBT-equivalent linewidth of $46.74\pm0.54$ MHz. A narrow-linewidth coherent laser provides the reference field. Fig.~\ref{fig:setup} shows the pump preparation and overall optical arrangement. Experimental parameters are given in Methods.

A microring resonator provides a natural platform for observing this statistical transformation. Resonant enhancement produces photon pairs with high spectral brightness. The narrow cavity modes also provide a photon lifetime long enough for the temporal correlations to be resolved. This photon response introduces a second timescale that shapes the measured HBT correlation together with the pump coherence time. The SiN microring used here has a measured resonance linewidth of approximately $125$ MHz and a free spectral range of about $100$ GHz, as shown in Fig.~\ref{fig:setup}. Residual pump light is removed after SFWM. The non-degenerate signal and idler photons are then separated for correlation measurements.

Fig.~\ref{fig:ring}a compares the incident-field statistics. The filtered ASE shows large temporal fluctuations and a broad, asymmetric intensity distribution, in clear contrast to the narrow distribution of the coherent reference. The fitted thermal intensity distribution includes the electronic readout noise and is used mainly to visualize the pump fluctuations. Quantitative characterization is provided by the HBT measurement. Fig.~\ref{fig:ring}b shows $g_{u}^{(2)}(0)=1.78\pm0.01$ for the ASE and $g_{u}^{(2)}(0)=1.00\pm0.01$ for the coherent laser. The corresponding effective pump mode number is $M_p=1.28\pm0.01$, which sets the pump-statistics parameter used in the calculation.

The signal--idler correlation in Fig.~\ref{fig:ring}d is dominated by a single temporal peak under coherent pumping, whereas a second temporal component becomes resolved under ASE pumping. The broader ASE spectrum and variations in its detuning from the microring may change the relative excitation of unresolved resonant contributions; the physical interpretation is discussed in Supplementary Materials. We describe the observed trace with a two-component fit. Treating these components as mutually incoherent detected contributions, their integrated weights give an effective mode number of $M_q=1.31\pm0.07$ for the HBT model.

\begin{table*}[t]
\centering
\caption{Comparison of the microring HBT measurement with the effective-model calculations. Uncertainties are one standard deviation. The experimental row lists the measured incident-pump linewidth, while the calculation rows list the pump linewidth used in each run. “Rel. dev.” denotes the relative deviation between the calculated and measured value. Relative deviations are calculated from the unrounded values, whereas the quantities displayed in the table are rounded.}
\label{tab:ring-comparison}
\small
\setlength{\tabcolsep}{4.5pt}
\begin{tabular*}{\textwidth}{@{\extracolsep{\fill}}lccccc@{}}
\toprule
Result & Pump linewidth (MHz) & $g_{u}^{(2)}(0)$ & Rel. dev. (\%) & HBT linewidth (MHz) & Rel. dev. (\%) \\
\midrule
Experiment & $46.74\pm0.54$ & $7.58\pm0.04$ & \textemdash & $58.98\pm0.39$ & \textemdash \\
Incident pump calculation & $46.74\pm0.54$ & $7.87\pm0.21$ & $3.64$ & $66.40\pm1.55$ & $11.17$ \\
Effective filtered pump calculation & $40.27\pm0.45$ & $7.94\pm0.22$ & $4.49$ & $59.17\pm1.44$ & $0.33$ \\
\bottomrule
\end{tabular*}
\end{table*}

The pump fluctuations produce a pronounced change in the photon statistics. Under coherent pumping, the intensity correlation reaches $g_{u}^{(2)}(0)=2.01\pm0.02$, close to the value expected for spontaneous pair generation into an approximately single optical mode. Fig.~\ref{fig:ring}c shows that replacing the coherent field with ASE increases the correlation to $g_{u}^{(2)}(0)=7.58\pm0.04$. Because the SFWM generation rate is weighted by $I_p^2$, strong pump excursions generate disproportionately large bursts of photon pairs. These rate fluctuations add to the bosonic bunching of the generated photons and produce the observed super-bunched statistics. A pump correlation of 1.78 thus produces an unconditional photon correlation exceeding 7.5.

The temporal correlation changes just as strongly. Under coherent pumping, the HBT trace gives a linewidth of $141.50\pm4.15$ MHz, which serves as the experimental reference for the photon response timescale. Under ASE pumping, the effective HBT linewidth decreases to $58.98\pm0.39$ MHz. This corresponds to an approximately 2.4-fold increase in correlation time. The ASE-pumped HBT trace combines the bunching of the detected optical mode with slower fluctuations of the pair generation rate inherited from the pump. Together, these contributions extend the second-order correlation beyond the photon response timescale. The narrower effective HBT linewidth therefore reflects an additional pump-dependent timescale, rather than an equivalent narrowing of the optical photon spectrum.

For quantitative comparison, the effective microring model describes the fluctuating pump using the measured $M_p$ and pump HBT linewidth. The photon response is calibrated from the coherent-pump HBT measurement, while $M_q$ describes the detected mode mixture. Using the incident ASE linewidth of $46.74\pm0.54$ MHz gives $g_{u}^{(2)}(0)=7.87\pm0.21$ and an effective HBT linewidth of $66.40\pm1.55$ MHz. The corresponding relative deviations are $3.64\%$ and $11.17\%$. The microring also filters the pump spectrum coupled into the resonator. We include this effect through the spectral overlap treatment described in Supplementary Materials, which gives an effective pump linewidth of $40.27\pm0.45$ MHz. The resulting predictions are $g_{u}^{(2)}(0)=7.94\pm0.22$ and an effective HBT linewidth of $59.17\pm1.44$ MHz, with relative deviations of $4.49\%$ and $0.33\%$, respectively (Table~\ref{tab:ring-comparison}). For this effective filtered-pump calculation, both observables agree with the measurements within two combined standard deviations. This agreement supports the statistical model and the treatment of the relevant temporal scales.

Super-bunching also increases the relative weight of multipair events and can raise the conditional second-order correlation of a heralded photon. The magnitude of this effect depends strongly on the mean pair generation probability. Figs.~\ref{fig:ring}e,f show the heralded HBT measurement. At low heralding rate, the correlation reaches $g_H^{(2)}(0)\simeq0.04$. Pronounced single-photon antibunching is therefore preserved despite the strong unconditional bunching. As the heralding rate increases, $g_H^{(2)}(0)$ rises because multipair generation becomes more important. Fluctuation-amplified photon statistics remain compatible with high-quality heralded single-photon generation when the source is operated at sufficiently low pair probability.

\begin{figure*}[t]
    \centering
    \safeincludegraphics[width=0.97\textwidth]{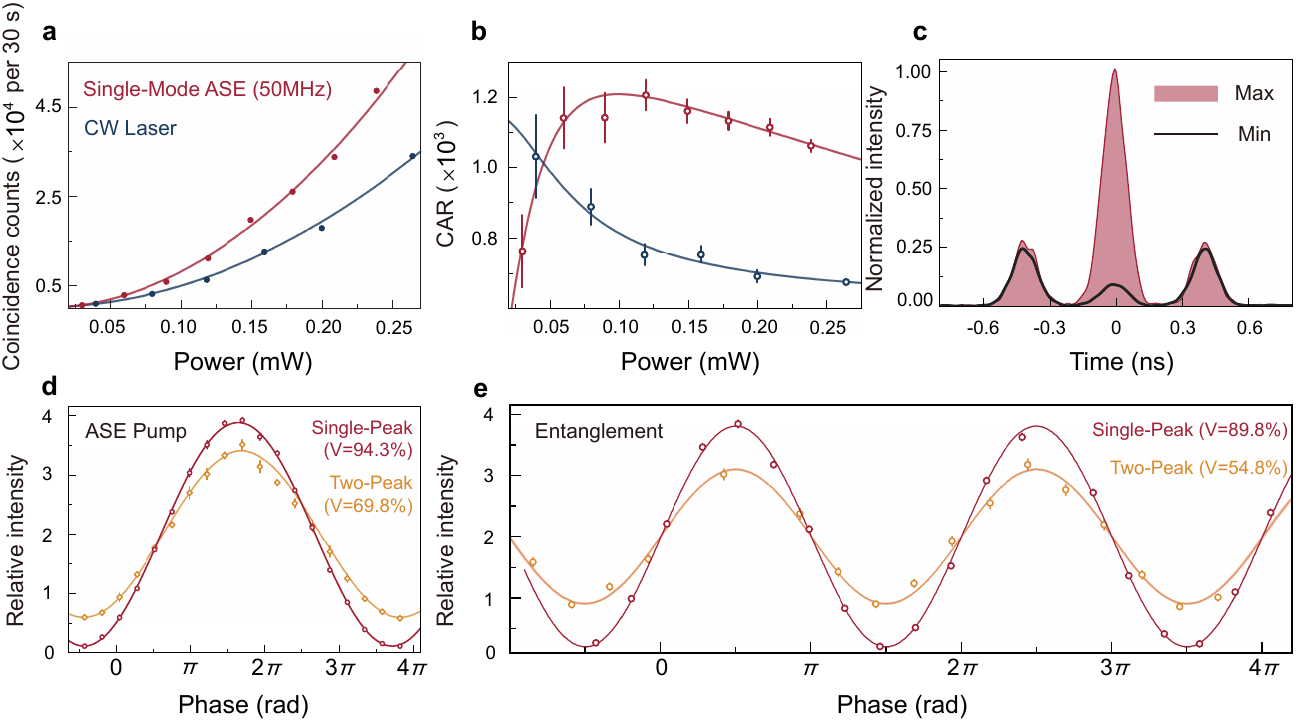}
    \caption{\textbf{Fluctuation-enhanced pair generation and time--energy entanglement.}
    \textbf{a,} Coincidence counts as a function of average pump power for single-peak filtered ASE ($g_{u}^{(2)}(0)=1.78$) and the coherent reference. Solid curves are quadratic fits.
    \textbf{b,} Coincidence-to-accidental ratio for the two pump fields.
    \textbf{c,} Representative coincidence histograms at the maximum and minimum of the two-photon interference fringe.
    \textbf{d,} Interference of the filtered ASE pump under single-peak and two-peak FP-filtering conditions, with fitted visibilities of $(94.27\pm0.30)\%$ and $(69.84\pm0.41)\%$, respectively.
    \textbf{e,} Corresponding two-photon interference fringes, with visibilities of $(89.84\pm0.40)\%$ and $(54.79\pm1.04)\%$. Solid curves are sinusoidal fits.}
    \label{fig:waveguide}
\end{figure*}

\subsection*{Quantum state generation with ASE pumping}

The microring measurements establish that pump fluctuations can strongly reshape the photon statistics while preserving pronounced heralded antibunching. We next examine whether the same ASE pump can also support the spectral and temporal coherence required for entangled photon-pair generation. For a particular temporal pump field $\alpha_{p,\xi}(t)$, labelled by $\xi$, the low-gain biphoton state can be written as

\begin{equation}
\begin{aligned}
|\psi_\xi\rangle
&\simeq
|0\rangle+
\eta
\iint
\Phi(\omega_s,\omega_i)\,
\mathcal P_\xi(\Omega_\Sigma)\\
&\quad\times
\hat a_s^\dagger(\omega_s)
\hat a_i^\dagger(\omega_i)
|0\rangle\,
\mathrm d\omega_s\,\mathrm d\omega_i,\\
\mathcal P_\xi(\Omega_\Sigma)
&=
\int
\widetilde{\alpha}_{p,\xi}(\Omega)
\widetilde{\alpha}_{p,\xi}(\Omega_\Sigma-\Omega)
\,\mathrm d\Omega .
\end{aligned}
\label{eq:biphoton-main}
\end{equation}

Here, $\eta$ is the low-gain pair generation amplitude and $\Omega_\Sigma=\omega_s+\omega_i-2\omega_p$. Here and in Supplementary Materials, $\omega_p$ denotes the pump centre frequency for the measurement under consideration. The function $\Phi(\omega_s,\omega_i)$ represents the deterministic spectral response of the waveguide, including phase matching over the finite interaction length. The term $\mathcal P_\xi$ is the self-convolution of the pump spectrum. The fluctuating phase of ASE is contained in the spectral amplitudes $\widetilde{\alpha}_{p,\xi}$. In Eq.~\eqref{eq:biphoton-main}, these phases enter through the products $\widetilde{\alpha}_{p,\xi}(\Omega)\widetilde{\alpha}_{p,\xi}(\Omega_\Sigma-\Omega)$. Each product is the pair generation amplitude associated with two pump-frequency components whose sum fixes $\Omega_\Sigma$. The observed two-photon state is obtained by averaging the corresponding density operators over the pump fluctuations, $\hat\rho_{si}\propto\langle|\psi_\xi^{(2)}\rangle\langle\psi_\xi^{(2)}|\rangle_\xi$, where $|\psi_\xi^{(2)}\rangle$ denotes the two-photon component of Eq.~\eqref{eq:biphoton-main}.

For a zero-mean Gaussian ASE field, this averaging removes coherence between components with different total pair frequencies. Within a given total-frequency component, the signal--idler amplitudes share the same pump-pair factor and remain coherently superposed through the phase-matching response. The absence of a deterministic optical phase in ASE therefore does not preclude the spectral coherence required for frequency entanglement. A full density-matrix treatment is given in Supplementary Materials.

We further examine how the fluctuating pump affects the pair generation rate and the coherence relevant to time--energy entanglement. The mean SFWM rate follows $R\propto\langle I_p^2\rangle=g_{u}^{(2)}(0)\langle I_p\rangle^2$, so thermal-like intensity fluctuations enhance the average pair generation probability at the same mean pump power. Time--energy entanglement is probed through interference between indistinguishable pair creation amplitudes separated by an interferometer delay $\tau$.\cite{franson1989} The normalized coherence of these amplitudes and the corresponding Franson visibility are

\begin{equation}
\begin{aligned}
\gamma_{\mathcal A}(\tau)
&=
\frac{
\left\langle
\alpha_p^{*2}(t)\alpha_p^2(t+\tau)
\right\rangle
}{
\left\langle|\alpha_p|^4\right\rangle
}
=
\left[g_p^{(1)}(\tau)\right]^2,\\
V_{\mathrm F}(\tau)
&=
\mathcal V_{\mathrm{inst}}
\left|g_p^{(1)}(\tau)\right|^2 .
\end{aligned}
\label{eq:franson-main}
\end{equation}

For Gaussian ASE, Wick factorization gives the relations in Eq.~\eqref{eq:franson-main}. The factor $\mathcal V_{\mathrm{inst}}$ is the visibility expected for fully coherent pair amplitudes in the same interferometer. It accounts for the finite contrast of the measurement without changing the dependence on pump coherence. The intensity fluctuations enhance the mean pair rate but do not add an independent reduction of the normalized two-photon coherence. The detailed derivation is given in Supplementary Materials.

We test these properties in a complementary experiment using a $1$-cm silicon strip waveguide with a $450\times220~\mathrm{nm}^2$ cross section and an insertion loss of approximately $4.3$ dB, as shown in Fig.~\ref{fig:setup}. The broad SFWM spectrum gives the generated photons a much shorter coherence time than the narrowband ASE pump. This provides the temporal hierarchy required for Franson interference. The same condition is less favorable in the microring because the pump and photon linewidths are comparable. Without resonant pump coupling, the straight waveguide also allows a more direct comparison at the same average pump power. To probe time--energy entanglement, the photon pairs pass through a common unbalanced Mach--Zehnder interferometer with a relative delay of approximately $400$ ps. The signal and idler photons are spectrally separated after the interferometer.

Fig.~\ref{fig:waveguide}a shows the pair generation rate for coherent pumping and for filtered ASE with one FP transmission peak, characterized by $g_{u}^{(2)}(0)=1.78$. Both pumps exhibit the expected quadratic dependence on average power, with a fitted generation-coefficient ratio of approximately $1.72$. Its proximity to the independently measured pump bunching supports the interpretation that intensity fluctuations enhance the mean SFWM rate through the second intensity moment. Under ASE pumping, the maximum measured coincidence-to-accidental ratio exceeds $1.1\times10^3$, as shown in Fig.~\ref{fig:waveguide}b.

We finally examine the temporal coherence and time--energy entanglement of the generated pairs. With one FP transmission peak, the pump interference visibility is $(94.27\pm0.30\%)$, as shown in Fig.~\ref{fig:waveguide}d. When two neighbouring FP transmission peaks contribute, the visibility decreases to $(69.84\pm0.41\%)$ at the same interferometer delay. The two spectral contributions acquire different phases across the delay and partially cancel in the first-order interference. Fig.~\ref{fig:waveguide}e shows the corresponding two-photon visibility, which changes from $(89.84\pm0.40\%)$ to $(54.79\pm1.04\%)$. This decrease is qualitatively consistent with the pump-coherence dependence in Eq.~\eqref{eq:franson-main}. The high Franson visibility in the single-peak condition provides evidence of time--energy entanglement. Together, the rate and interference measurements show that strong pump-intensity fluctuations can enhance photon-pair generation without imposing a corresponding loss of coherence between alternative pair creation times.

\section*{Discussion}\label{sec:discussion}

This work demonstrates fluctuation-driven amplification of quantum statistics in a spontaneous multiphoton process. Our theoretical formulation extends spontaneous quantum generation beyond the coherent-pump limit and shows how nonlinear intensity weighting combines with bosonic bunching to produce pronounced super-bunching. The finite cavity response governs the transfer of pump fluctuations to the photon flux, while pump coherence introduces an additional temporal scale. This interplay accounts for both the enhanced HBT peak and the extended correlation time observed in the microring, establishing a direct connection between driving-field statistics and the strength and temporal structure of the generated correlations. The heralded measurements further show that strong unconditional bunching can coexist with single-photon antibunching at low generation rates. Complementary time--energy entanglement demonstrates that the same fluctuating pump can support coherent superpositions of pair-creation times. Together, these observations show that pronounced statistical amplification can coexist with key quantum-state properties.

Beyond spontaneous quantum generation, the same physical mechanism can be extended to other multiphoton interactions. Second-harmonic generation, for example, offers a promising route towards bright single-mode light with enhanced bunching. Such sources could be particularly valuable for remote optical sensing,  where pronounced intensity correlations must be combined with useful detection rates after propagation loss. More generally, our findings show that photon statistics can be reshaped through nonlinear interactions rather than remaining fixed by the source. This principle offers a blueprint for developing optical sources with tailored photon statistics, laying the groundwork for bringing advanced statistical control of light in optical measurements and nonlinear light--matter interactions.

\section*{Methods}\label{sec:methods}
\setcounter{equation}{0}
\renewcommand{\theequation}{M\arabic{equation}}

\subsection*{Theory of fluctuation-driven SFWM in a microring}

The filtered ASE pump is described as a stationary fluctuating complex field with a slowly varying envelope $\alpha_p(t)$. Its spectral amplitudes contain both the amplitude and phase fluctuations relevant to the nonlinear interaction,

\begin{equation}
\begin{aligned}
\alpha_p(t)
&=
\frac{1}{\sqrt{2\pi}}
\int
\widetilde{\alpha}_p(\Omega)
e^{-i\Omega t}\,
\mathrm d\Omega,\\
\left\langle
\widetilde{\alpha}_p^*(\Omega)
\widetilde{\alpha}_p(\Omega')
\right\rangle
&=
S_p(\Omega)\delta(\Omega-\Omega').
\end{aligned}
\label{eq:M1}
\end{equation}

Here, $S_p(\Omega)$ is the pump power spectrum. A finite spectral width gives a finite first-order coherence time. The vanishing mean field is consistent with the thermal-like intensity fluctuations measured experimentally. For a Lorentzian-equivalent pump spectrum of full width at half maximum (FWHM) of $\nu_P$, the normalized first-order coherence is $g_p^{(1)}(\tau)=\exp(-\pi\nu_P|\tau|)$. The statistical description of the filtered ASE and its behaviour under linear amplification are developed further in Supplementary Materials.

Using the SFWM interaction in Eq.~\eqref{eq:sfwm-hamiltonian}, the signal and idler fields in the microring obey the linearized Heisenberg--Langevin equations\cite{vernon2015strong}

\begin{equation}
\begin{aligned}
\dot{\hat a}_s(t)
&=
-\left(\frac{\kappa_s}{2}+i\Delta_s\right)\hat a_s(t)
+\mathcal G(t)\hat a_i^\dagger(t)
+\sum_\ell\sqrt{\kappa_{s\ell}}\,
\hat b_{s\ell,\mathrm{in}}(t),\\
\dot{\hat a}_i^\dagger(t)
&=
-\left(\frac{\kappa_i}{2}-i\Delta_i\right)\hat a_i^\dagger(t)
+\mathcal G^*(t)\hat a_s(t)
+\sum_\ell\sqrt{\kappa_{i\ell}}\,
\hat b_{i\ell,\mathrm{in}}^\dagger(t),
\end{aligned}
\label{eq:M2}
\end{equation}

Here, $\kappa_\mu$ is the total decay rate of mode $\mu=s,i$, and $\Delta_\mu$ is its detuning. The operator $\hat b_{\mu\ell,\mathrm{in}}$ denotes the vacuum input associated with coupling to the bus waveguide or intrinsic loss. The nonlinear coupling $\mathcal G(t)=g\alpha_p^2(t)$ transfers the temporal pump fluctuations directly to the pair creation amplitude.

In the spontaneous low-gain limit, the cavity evolution can be treated perturbatively. The causal cavity-amplitude response is $h_\mu(t)=\exp[-(\kappa_\mu/2+i\Delta_\mu)t]\Theta(t)$: the amplitude decays at rate $\kappa_\mu/2$, the detuning produces phase evolution, and the Heaviside function enforces causality.\cite{vernon2015lossy} The contribution of the pump over the signal and idler response times is described by

\begin{equation}
\mathcal K(t,v;\mathcal G)
=
\int_v^t
h_s(t-u)\,
\mathcal G(u)\,
h_i^*(u-v)\,
\mathrm du .
\label{eq:M3}
\end{equation}

The pump amplitude at $u$ is weighted by the idler response from $v$ to $u$ and by the signal response from $u$ to the observation time $t$. Slowly varying fluctuations are transferred efficiently to the generated photon flux. Faster fluctuations are averaged more strongly by the cavity response. The perturbative solution is derived in Supplementary Materials.

For a given pump field $\alpha_p(t)$, the low-gain signal field is Gaussian. We denote its first-order covariance by $C_{\mathcal G}(t,t')$ and define the photon flux as $N_{\mathcal G}(t)=C_{\mathcal G}(t,t)$. In the input--output description used here, the covariance is

\begin{equation}
\begin{aligned}
C_{\mathcal G}(t,t')
&=
\kappa_{se}
\sum_\ell\kappa_{i\ell}
\int_{-\infty}^{\min(t,t')}
\mathcal K_\ell^*(t,v;\mathcal G)\\
&\qquad\times
\mathcal K_\ell(t',v;\mathcal G)\,
\mathrm dv .
\end{aligned}
\label{eq:M4}
\end{equation}

Averaging the Gaussian fourth-order correlation over the fluctuating pump then gives the unconditional second-order correlation

\begin{equation}
\begin{aligned}
g_{u}^{(2)}(t,t')
=
\frac{
\left\langle
N_{\mathcal G}(t)N_{\mathcal G}(t')
\right\rangle
+
\left\langle
|C_{\mathcal G}(t,t')|^2
\right\rangle
}{
\left\langle N_{\mathcal G}(t)\right\rangle
\left\langle N_{\mathcal G}(t')\right\rangle
}.
\end{aligned}
\label{eq:M5}
\end{equation}

The first term describes correlations of the pair generation rate after the cavity response. The second is the bosonic exchange term from Gaussian moment factorization. Its magnitude is set by the temporal overlap of the detected optical mode. Equation~\eqref{eq:M5} forms the basis of the numerical calculation used for the microring comparison.

For comparison with experiment, the pump is specified by its HBT linewidth and effective mode number $M_p=[g_u^{(2)}(0)-1]^{-1}$. We model it using two statistically independent thermal modes whose relative mean intensities reproduce the measured $M_p$. Their pair-creation amplitudes account for annihilating two photons from either pump mode or one photon from each mode, as defined in Supplementary Materials. The relative detected-mode weights $p_q$ are obtained from the signal–idler correlation and define $K_q=\sum_q p_q^2=1/M_q$. For the present experiment, $M_q=1.31\pm0.07$. The multimode correlation and numerical evaluation are given in Supplementary Materials.

The parameter scans in Fig.~\ref{fig:theory} vary the pump bandwidth or fluctuation strength while the remaining quantities are held constant. For the comparison in Fig.~\ref{fig:ring}, the pump mode number and linewidth are obtained from the pump HBT measurement. The photon response is calibrated under coherent pumping, and the optical-mode weights are taken from the signal--idler correlation. These measured quantities are used before calculating the ASE-pumped HBT trace. Experimental calibration, numerical implementation and uncertainty propagation are described below. Analytical limits are given in Supplementary Materials.

Strong unconditional bunching also modifies the multipair contribution relevant to heralded single-photon operation. Let $\bar\mu\ll1$ denote the mean number of generated pairs within an effective heralding window. The factor $K_q=\sum_q p_q^2$ describes the mode distribution of the generated photons. Pump-driven rate fluctuations increase the second factorial moment of the pair number. For this analytical estimate, the detected modes share a common pump-driven rate modulation with fixed relative occupations. We write

\begin{equation}
\begin{aligned}
C_\mu
&=
\frac{\langle\mu_\xi^2\rangle_\xi}{\bar\mu^2},
\qquad
G=(1+K_q)C_\mu .
\end{aligned}
\label{eq:M6}
\end{equation}

In the low-gain limit, the one-pair probability scales as $P_1\simeq\bar\mu$, while the two-pair probability is $P_2\simeq G\bar\mu^2/2$. For a non-number-resolving herald detector with efficiency $\eta_h$, the corresponding heralded correlation is

\begin{equation}
g_H^{(2)}(0)
\simeq
(2-\eta_h)G\bar\mu .
\label{eq:M7}
\end{equation}

Pump-induced super-bunching increases the multipair contribution at a given mean pair probability. Even so, $g_H^{(2)}(0)$ remains proportional to the source brightness per heralding window and approaches zero in the low-gain limit. With unchanged detection settings, the heralding count rate provides an experimental measure of this brightness, consistent with Fig.~\ref{fig:ring}f. The derivation is given in Supplementary Materials.

\subsection*{Experimental setup and measurements}

The experimental setup is shown in Fig.~\ref{fig:setup}. The fluctuating pump is prepared from an amplified spontaneous emission source and spectrally selected near 1550 nm. A 0.3-nm fibre Bragg grating first restricts the ASE bandwidth. After polarization adjustment, an erbium-doped fibre amplifier increases the available spectral power density. The field then passes through two free-space Fabry--Pérot (FP) filters with bandwidths of approximately 5 GHz to suppress out-of-band transmission. A final narrowband FP cavity selects the pump spectrum used for SFWM. Its free spectral range is approximately 6.91 GHz. A direct frequency scan gives a linewidth of 47.26 MHz, while the HBT-derived value used in the statistical analysis is $46.74\pm0.54$ MHz. The FP resonance is temperature stabilized during the measurements. The effect of linear optical amplification on the ASE statistics is derived in Supplementary Materials. A tunable coherent laser (Toptica DL100, MHz-scale linewidth) operating near 1550 nm is used as the reference pump and for resonator characterization. For the microring measurements, the pump frequencies are adjusted to address the selected cavity resonance.

For the resonant measurements, the pump is coupled to a SiN microring with a radius of $R=230\,\mu\mathrm{m}$ and a waveguide cross section of $1.2\times0.8\,\mu\mathrm{m}^2$. The spacing between the ring and the waveguide is 800 $\mathrm{nm}$ and the total insertion loss is approximately 5 dB. The microring has a free spectral range of approximately 100 GHz and a scanned resonance linewidth of 124.75 MHz, obtained from an asymmetric fit. Possible contributions to the asymmetric line shape include thermo-optic drift during the scan and unresolved backscattering-induced coupling between counter-propagating cavity modes.\cite{carmon2004thermal,li2016backscattering}  Both the microring and the narrowband FP filter are temperature stabilized. After SFWM, residual pump light is removed by band-stop filtering. A dense wavelength-division multiplexer separates the non-degenerate signal and idler photons into channels C32 and C36 before detection. The same optical routing is used for coherent and ASE pumping so that both measurements address the same signal and idler resonances.

The pump and photon statistics are measured in Hanbury Brown--Twiss configurations using fibre beam splitters and superconducting nanowire single-photon detectors. A time-to-digital converter (TDC, TimeHarp 260, PicoQuant) records the photon arrival times used to construct the second-order correlation histograms. Signal--idler correlations are recorded with the same system to characterize the optical modes participating in pair generation. For the heralded measurement, the signal photon serves as the herald. The idler photon is divided by a 50:50 fibre beam splitter, and each output is detected by a superconducting nanowire detector. A three-channel coincidence unit (UQD Logic16) records the corresponding two-fold and three-fold coincidences used to determine $g_H^{(2)}(0)$. The correlation analysis and mode-number extraction are described in Data analysis and numerical modelling.

For the complementary measurements of pair generation rate and time--energy coherence, the prepared pump is directed to a 1-cm silicon strip waveguide with a $450\times220~\mathrm{nm}^2$ cross section and a total insertion loss of approximately 4.3 dB. The coherent and ASE measurements use the same optical path and the same pump-power reference, measured with an integrating-sphere photodiode power sensor (Thorlabs S145C). Residual pump light is suppressed after the waveguide.

To probe time--energy entanglement, the generated photon pairs are sent through a common unbalanced Mach--Zehnder interferometer (MZI) with a relative delay of approximately 400 ps. Pair creation amplitudes separated by a delay $\tau$ have the normalized coherence $\gamma_{\mathcal A}(\tau)=\langle\alpha_p^{*2}(t)\alpha_p^2(t+\tau)\rangle/\langle|\alpha_p|^4\rangle$. For the Gaussian pump model, $\gamma_{\mathcal A}(\tau)=[g_p^{(1)}(\tau)]^2$, as derived in Supplementary Materials. A polarization controller before the MZI compensates its polarization sensitivity and is adjusted to obtain an approximately 1:1 transmission-to-reflection ratio. The interferometer phase is controlled thermally with millikelvin-level stability. After the MZI, a 200-GHz DWDM separates the signal and idler into channels C20 and C48, with nominal channel centres at 1559.79 nm and 1538.98 nm, respectively. The TDC records the time-tagged events used to obtain coincidences and accidentals.

For the single-peak measurements, the final FP filter is adjusted so that one transmission mode dominates the ASE spectrum. This condition gives the pump statistics used for the rate measurements, with $g_{u}^{(2)}(0)=1.78\pm0.01$. A second condition is obtained by tuning the FP cavity so that two neighbouring transmission modes contribute. Slow variations of the FP resonance produce a small imbalance between the two transmitted components. Pump interference through the MZI is measured independently to characterize the first-order coherence. Two-photon interference is then recorded under the same filtering condition. The central coincidence peak contains the short--short and long--long amplitudes that form the Franson fringe. The two side peaks correspond to the distinguishable short--long and long--short paths. All visibilities reported in Fig.~\ref{fig:waveguide} are raw values without accidental subtraction.

\subsection*{Data analysis and numerical modelling}

The HBT measurements are analysed directly from the recorded coincidence histograms. Before detector timing broadening, the correlation is described by a symmetric exponential excess on a constant background. The fit includes convolution with a Gaussian timing response. The fitted decay time $\tau_c$ defines the Lorentzian-equivalent HBT linewidth, $\nu_{\mathrm{HBT}}=(2\pi\tau_c)^{-1}$. The zero-delay correlation is obtained from the deconvolved peak-to-background ratio. For ASE-pumped photons, the reported width is termed an effective HBT linewidth because the correlation contains both pump-driven rate fluctuations and the photon exchange contribution. The comparison uses the deconvolved $g_u^{(2)}(0)$ and HBT linewidth from this fit. No additional timing convolution is applied to the optical calculation.

The temporal components in the signal--idler cross-correlation are fitted by two asymmetric exponential peaks with a common background and the detector timing response. Their relative mode weights are obtained from the integrated areas $S_j=A_j(\tau_{r,j}+\tau_{f,j})$. With $p_j=S_j/(S_1+S_2)$, the effective mode number of the generated photons is

\begin{equation}
M_q=\frac{1}{p_1^2+p_2^2}.
\label{eq:M8}
\end{equation}

The uncertainty of $M_q$ is evaluated by Monte Carlo propagation of the covariance matrix from the nonlinear fit. Correlated parameter sets are sampled from the fitted distribution and converted to peak areas, mode weights and $M_q$. The standard deviation of the resulting distribution is reported as the one-standard-deviation uncertainty, giving $M_q=1.31\pm0.07$.

The microring calculation uses parameters obtained from the corresponding measurements. The pump is described by $M_p$ and its HBT linewidth. The photon response is calibrated against the coherent-pump HBT linewidth, while $M_q$ parameterizes the assumed detected mode mixture. The filtered-pump calculation uses an effective linewidth of $40.27\pm0.45$ MHz obtained from the spectral-overlap approximation described in Supplementary Materials.

The numerical calculation follows the low-gain model developed above. The two independent thermal pump modes share a common correlation time, and their relative mean intensities reproduce the measured $M_p$. The pair-creation amplitudes defined in Supplementary Materials are propagated through the calibrated signal and idler responses. The detected intensities are added, with the measured factor $K_q$ weighting the bosonic exchange term. The parameter scans use stationary and two-time moment equations, while the comparison with experiment uses stochastic propagation of the same model. The theoretical peak is evaluated at zero delay, and the effective HBT linewidth is obtained by fitting the correlation excess to a single exponential. Input uncertainties are propagated by finite differences and combined with the Monte Carlo sampling uncertainty for the experimental comparison.

The heralded correlation is obtained from the three-fold coincidence events using the same 5-ns coincidence window throughout each measurement series. Its dependence on heralding count rate is interpreted using the low-gain pair-number model derived in Supplementary Materials.

For the straight waveguide measurements, the pair generation rate is obtained from the quadratic dependence of the coincidence counts on average pump power, and the fitted coefficients are used to compare ASE and coherent pumping. Coincidence-to-accidental ratios are calculated using coincidence and displaced accidental windows acquired under the same conditions.

The pump and two-photon interference fringes are fitted on the original count scale using $C(x)=C_0[1+V\sin(\pi(x-x_c)/w)]$, where $V$ is the interference visibility. Its uncertainty is evaluated by Monte Carlo resampling of the measured data. Each point is sampled according to its experimental uncertainty, and the full fringe is refitted for every synthetic data set. The standard deviation of the resulting visibility distribution is quoted as the one-standard-deviation uncertainty. All reported visibilities are obtained without accidental subtraction. Normalization is applied only for graphical presentation.

\section*{Data availability}
All data that support the plots within this paper and other findings of this study are available from the corresponding authors upon reasonable request.

\section*{Acknowledgments}
This work is supported by the National Key Research and Development Program of China (2022YFB3903102 and 2022YFB3607700), the National Natural Science Foundation of China (NSFC; 62435018), the Quantum Science and Technology National Science and Technology Major Project (2024ZD0300800, 2023ZD0300800), the Innovation Program for Quantum Science and Technology (2021ZD0301100), USTC Research Funds of the Double First-Class Initiative (YD2030002023), and the Research Cooperation Fund of SAST, CASC (SAST2022-075). This work was partially carried out at the USTC Center for Micro and Nanoscale Research and Fabrication

\section*{Author contributions}

Y.W.S., Z.H.Z. and S.W. contributed equally to this work. C.W., Z.Y.Z. and B.S.S. conceived the project, supervised the work and acquired the funding. Y.W.S., Z.H.Z., S.W. and Y.H.L. conceived and designed the experiments. Y.W.S. developed the theory under the supervision of Z.Y.Z. and B.S.S. H.C.W., B.W.L. and J.P.L. analysed the data. S.W. and C.H.D. fabricated the device. Y.W.S., Z.H.Z. and H.C.W. performed the experiments. Y.W.S., Z.H.Z., Z.Y.Z., B.S.S. and G.C.G. wrote the manuscript. All authors contributed to discussions and the interpretation of the results.

\bibliographystyle{unsrtnat}
\bibliography{references}

\balance
\end{document}